# Fluidic Approach to Corrective Eyewear Manufacturing in Low-Resource Settings

Mor Elgarisi[1], Omer Luria[1], Yotam Katzman[1], Daniel Widerker[1], Valeri Frumkin[1,2], Moran Bercovici[1,*]

[1]Faculty of Mechanical Engineering, Technion – Israel Institute of Technology, Haifa, Israel.

[2]Current affiliation: Department of Mechanical Engineering, Boston University, Boston, MA, USA.

[*]Corresponding author: mberco@technion.ac.il

**Abstract**

Limited access to corrective eyewear remains a significant medical, societal, and economic challenge, even in the 21st century. More than 1 billion people suffer from uncorrected vision impairment, with the vast majority residing in developing countries. Decades of philanthropic efforts failed to supply even a small fraction of the demand, whereas local manufacturing using standard machining technologies remains out of reach due to inadequate resources. We here show that the Fluidic Shaping approach[1,2] can be utilized to create a new manufacturing modality for high-quality ophthalmic lenses that entirely eliminates the need for machining, and instead uses surface tension to shape liquid polymer volumes into prescription lenses. We present a compact device wherein a liquid photopolymer is injected into an elliptical frame submerged within an immersion liquid of equal density, resulting in two liquid surfaces whose minimum-energy states correspond to two lens surfaces. After several minutes of curing, a complete solid lens is obtained, requiring no post-processing. We provide an analytical model and experimental validation, showing that all spherical and cylindrical corrections can be attained by simply controlling the volume of the polymer and the frame's eccentricity. We demonstrate the fabrication of complete eyeglasses that meet and exceed industry standards, relying solely on a 1 gallon water container integrated with an array of low-power LEDs.

**Main**

Blindness is often perceived as a permanent medical condition, untreatable without highly advanced medical intervention. However, millions of blindness cases result from uncorrected refractive error – a condition that can be easily treated with prescription eyeglasses[3–6]. Over 90% of these cases are in the developing world[7], and remain untreated due to lack of access to corrective eyewear. This is true not only for blindness, but for all severity levels of vision impairment, making refractive error the leading untreated ophthalmic medical condition[4]. At present, 1 billion people in the world suffer from vision impairment and lack access to corrective eyewear[3–5,7,8].

Beyond the detrimental effects on the quality of life on the individual level[4,9–11], untreated vision impairment has substantial societal and economic consequences, including academic underperformance and reduced literacy[12–18], reduced workplace productivity[15,19–23], decrease in road safety[24–26], and gender inequality[3,27]. At present, these result in an annual global financial loss of $410 billion[4,28,29], which, due to population growth[30], is predicted to increase to $920 billion by 2050[31] – a $20 trillion loss over three decades.

While in the developed world optometry services, which include diagnosis, prescription, and lens fabrication, are widely accessible, they are scarce in the developing world[4,32]. For example, there are 221 optometrists per 1M people in high income countries (Europe, North America) but only 1 optometrist per 1M people in sub-Saharan Africa. While much work has been done in the past decade in developing accessible solutions for hdiagnosis and prescription[33–35], there remains a major challenge in providing individuals with appropriate eyeglasses.

The standard methods for fabrication of eyeglasses are based on mechanical processes such as machining and polishing, or molding[36–39]. These processes require significant energy and water resources, major capital investments in infrastructure, and operation by trained personnel. Such facilities are incompatible with low-resource settings, and thus attempts to satisfy the demand for eyeglasses in such regions have relied primarily on operating distribution logistics to deliver pre-fabricated lenses to the point of need[7]. To date, according to the World Economic Forum[5], all such initiatives were able to supply only a small fraction (0.3% in 2016) of the need. Given these circumstances, there is a pressing need to innovate lens manufacturing methodologies. For example, new technologies, which bring fabrication of high-quality eyeglasses close to the point-of-need, and that are appropriate for low-resource settings, have the potential of disrupting the existing modes of operation and providing a larger fraction of the population with access to corrective eyewear.

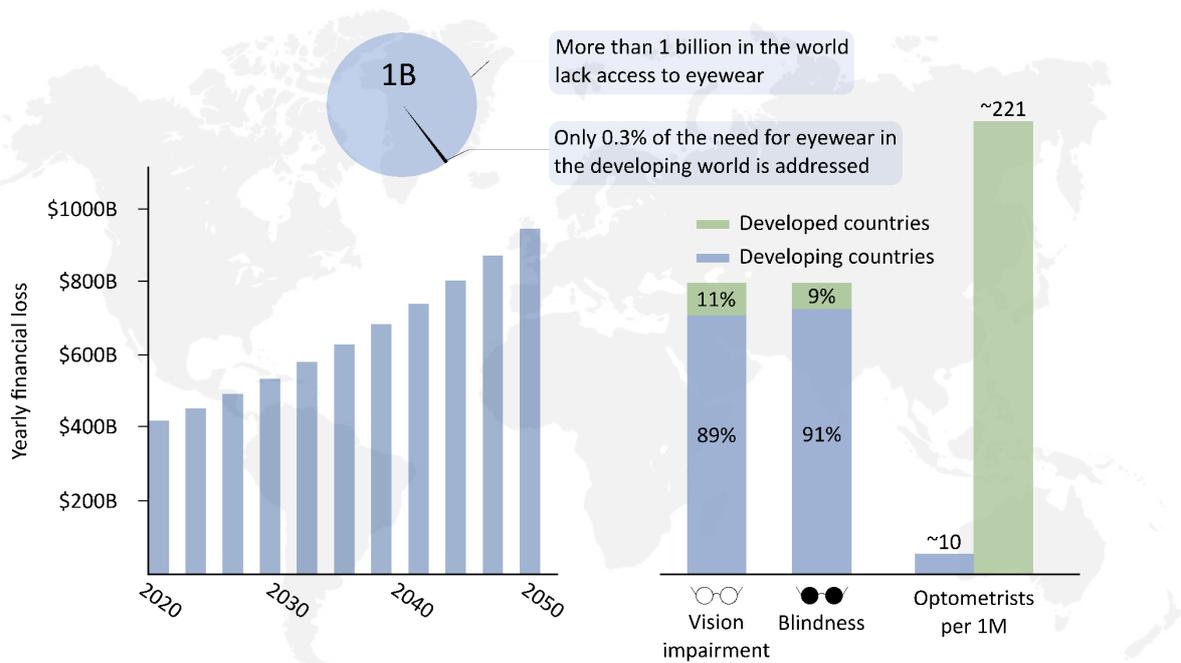

*Figure 1 – Eyewear inaccessibility presents global medical, social, and financial challenges* [Data collected from The Lancet[4], World Economic Forum[5], and World Health Organization[7]]. *Over a billion individuals, mostly in the developing world, require corrective eyewear, with merely 0.3% of that need currently addressed. This results in an estimated economic loss that exceeds $400 billion annually, and is projected to increase to over $920 billion by 2050. Optometry services are crucially lacking in developing countries, with ten optometrists per million people, as compared to 221 in developed countries. Notably, 89% of patients with visual impairment and 91% of those suffering from refractive-error blindness reside in low-resource areas that cannot support local ophthalmic manufacturing.*

Fluidic Shaping is a novel approach for fabrication of optical components that is based on shaping of liquid polymers using surface tension under neutral buoyancy conditions[2,40]. In this work, we leverage the Fluidic Shaping approach to create a complete solution for fabrication of prescription lenses that eliminates the need for any machining, polishing, or molding steps. Using a device whose footprint is 20 cm x 20 cm, with an energy consumption of 25 W, a pair of high-quality lenses (diopter variation of less than 0.15 D, and surface roughness of less than 1 nm) can be produced in under 10 minutes. As illustrated in Figure 2, the entire lens fabrication process takes place within a single device that transforms a volume of liquid polymer into a lens with any combination of spherical or cylindrical corrections.

The device consists of a container that is filled with an immersion liquid of density $\rho_{im}$, and is divided in two by a plate with a circular opening at its center. A frame, which will be used as the boundary for the injected lens, is placed into the opening and seals against it (Figure 2a). The inner surface of the frame has a bottom circular contour and an upper elliptical contour. The first step in creating the lens is injecting into the frame a volume $V_{lens}$ of a photopolymer with density $\rho_p$ that is immiscible with the immersion liquid (Figure 2b). At this step, an external channel that connects the two parts of the container remains open to allow for pressure equilibration. When setting the density of the immersion liquid to match that of the photopolymer, i.e., $\rho_p = \rho_{im}$, gravity forces are perfectly counter-balanced by buoyancy. Under these neutral buoyancy conditions, surface tension is the only force acting on the liquid, and will drive the system to minimize its surface area - corresponding to its minimum energy state. For example, in the basic case of a circular boundary, this minimum energy state corresponds to two identically spherical caps that can be convex or concave, depending on the volume $V_{lens}$ that was injected.

An additional degree of freedom in controlling the shape of the lens is achieved by closing the equilibration channel and adding immersion liquid to the bottom chamber, thus curving the lens upwards (Figure 2c). This results in a meniscus-shaped lens that provides physical clearance from the eye, and is the standard form of ophthalmic lenses today[41]. The final step is the solidification of the lens via exposure to UV light for several minutes (Figure 2d). The solid lens is then removed from the container and the fabrication process is complete – no further processing steps are required (Figure 2e). The system is then ready for re-use. Figures 2f and 2g show the setup and the liquid lens during solidification of a meniscus lens. Figure 2h shows the finished lens in its original frame, which once detached, may undergo any of the standard processes for fitting it into a desired frame. Fig. 2i presents several fully assembled eyeglasses with lenses fabricated by Fluidic Shaping.

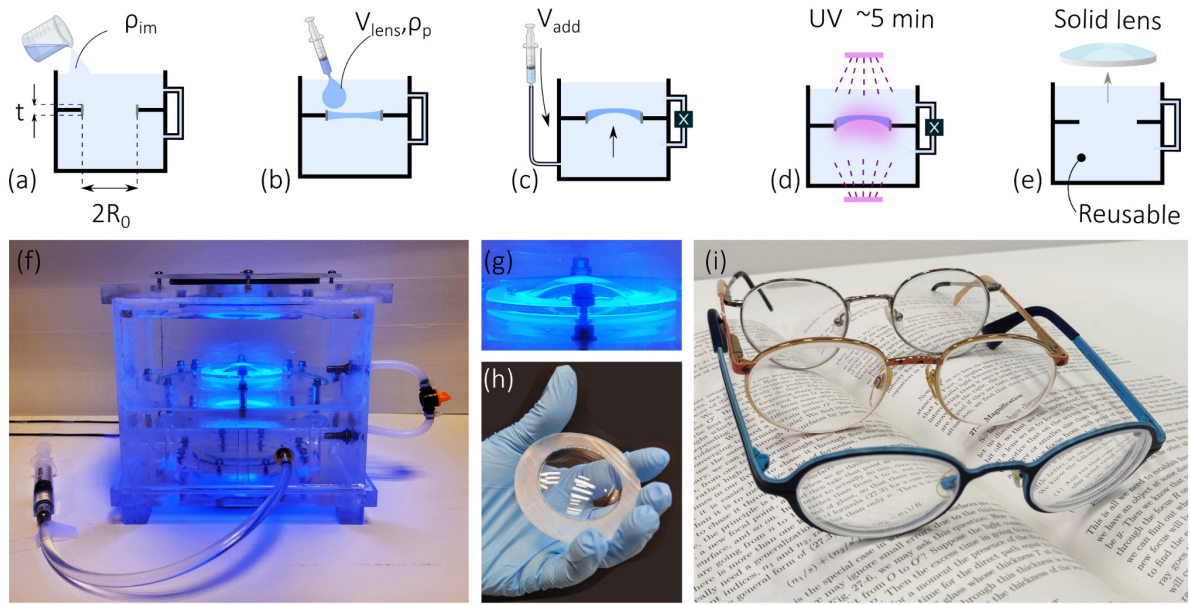

*Figure 2 – The Fluidic Shaping fabrication process, the experimental setup and fabricated lenses. (a-e): Schematic illustration of the method. (a) An aquarium is filled with an immersion liquid, that is immiscible with the lens polymer, and whose density is set to match that of the polymer. (b) A fixed volume of the curable liquid polymer, determined by the prescription, is injected into a bounding frame submerged within the immersion liquid. The shape of the liquid-liquid interface is determined by energy minimization, resulting in a symmetrical lens. (c) The lower part of the aquarium is injected with an additional immersion liquid, which pushes the lens upward and results in a meniscus lens. (d) Once the liquid shape is stable, it is polymerized using UV light. (e) The meniscus lens is now solid and can be taken out of the aquarium. The lens is ready to be used, with no further polishing or grinding processes needed. (f) The experimental setup, with a fluidic lens in it, under polymerization. (g) A zoom in view of the fluidic meniscus lens. (h) The lens after solidification, outside the aquarium, still in its frame. Once the lens is solidified, it can be taken out of its frame and placed on eyewear rims. (i) Fully assembled eyewear with lenses that were fabricated using the Fluidic Shaping method.*

An ophthalmic prescription for single vision lenses would typically consist of a spherical and a cylindrical correction[41,42]. The Fluidic Shaping method allows to achieve any combination of the two, by controlling two main parameters – the volume of the injected photopolymer, and the shape of the bounding frame. We proceed to describe an analytical model that relates the desired optical prescription to the geometrical parameters required for the fabrication. Figure 3(a) presents a schematic illustration of a fluidic lens formed within a frame of height $t$, consisting of a circular bottom contour of radius $R_0$ and an elliptical top contour with semi diameters $a,b$. The volume of the liquid photopolymer encompassed within the frame is $V_{lens}$, and the volume enclosed between its bottom surface and the x-y plane is $V_{enc}$. At steady state, the system achieves its minimum energy configuration, which can be described by the minimization of an energy functional for the top and bottom surface $h_{top}, h_{bot}$, as detailed in SI2. For the considered configuration, the bottom surface is guaranteed to be a spherical cap with a radius of curvature that we denote as $R_{bot}$, and the top surface is a toric surface with two radii of curvature, $R_{xtop}$ and $R_{ytop}$. The spherical power of such a lens is given by the 'lens maker's equation' [41],

$$P = (n-1)\left( \frac{1}{R_{top}} - \frac{1}{R_{bot}} + \frac{(n-1)d}{nR_{top}R_{bot}} \right), \quad (1)$$

where $R_{top} = \left(1/R_{xtop} + 1/R_{ytop}\right)^{-1}$ is the mean radius of curvature of the top surface, $d$ is the center thickness of the lens, and $n$ is the refractive index of the photopolymer. Similarly, the cylindrical power is given by

$$C = (n-1)\left( \frac{1}{R_{xtop}} - \frac{1}{R_{ytop}} \right)\left( 1 + \frac{(n-1)d}{nR_{bot}} \right). \quad (2)$$

Using the minimum energy model, we can relate the radii of curvature to the geometry of the frame and the injected volumes. Through the thin lens approximation[41] the spherical and cylindrical powers can be expressed as [See SI3]:

$$(3a) \quad P = 2(n-1)\left( \frac{(\Delta V_{lens} + V_{enc})(a^2 + b^2)}{\pi a^3 b^3} - \frac{h_0}{R_0^2} \right),$$

$$(3b) \quad C = 4(n-1)\frac{(\Delta V_{lens} + V_{enc})(a^2 - b^2)}{\pi a^3 b^3}, \quad (3)$$

where $\Delta V_{lens} = V_{lens} - \frac{\pi}{2}(a+b)R_0 t$, and $h_0 = \frac{2V_{enc}}{\pi R_0^2}$ is the clearance distance between the x-y plane and the center of the bottom surface (see Fig 3a).

Combining Equation 3a and 3b, the spherical and cylindrical powers can be related through

$$P = C \cdot \left( \frac{1}{e^2} - \frac{1}{2} \right) - 2(n-1)\frac{h_0}{R_0^2}, \quad (4)$$

indicating that any combination of spherical and cylindrical powers can be achieved by adjusting the eccentricity of the of frame, $e = \sqrt{1 - (b^2/a^2)}$.

Figure 3b shows the cylindrical power as a function of the spherical power and the eccentricity, for a lens with $R_0 = 25mm$, $n = 1.525$ and $h_0 = 3mm$. The figure demonstrates the degrees of freedom in the design, where for a lens of a given radius $R_0$ and a desired clearance $h_0$, one can select the desired spherical power, and obtain the required eccentricity in order to satisfy the desired cylindrical power. Using Equation 3, we can now retrieve the photopolymer volume, $\Delta V_{lens}$, required to produce the lens. SI4 presents a flowchart for determining the physical parameters (frame geometry and liquid volumes) required to achieve any spherical and cylindrical combination.

Figure 3c presents the required photopolymer volume as a function of the lens size, for the simplified case of zero cylindrical power. The displayed values encompass typical sizes for children's and adults' frames, but can naturally be extended beyond the presented range. The typical volume to be injected is on the order of milliliters. The blue curve in the figure shows the sensitivity of the diopter to inaccuracy in the injection. For example, for a 50 mm frame diameter, a 1/8 diopter deviation (which is well within the range that the human eye typically cannot discern[43]), results from a $50 \mu L$ inaccuracy in volume. Maintaining such a volume accuracy, on the order of 1% total volume, can be easily achieved with low-cost injection systems[44,45].

Contrary to the spherical power, which is dictated by the difference in curvature between the top and bottom surfaces and thus is weakly affected by $V_{enc}$, the cylindrical power is controlled only by the elliptic (top) surface and is thus much more sensitive to both $\Delta V_{lens}$ and $V_{enc}$. Figure 3d shows the cylindrical power of the lens as a function of eccentricity and the total volume $\Delta V_{lens} + V_{enc}$, for a base diameter of 50 mm and a refractive index 1.525. The blue curve describes the cylindrical power sensitivity to the combined injected volumes, which is an order of magnitude lower than that of the spherical power. The orange curve represents the cylindrical power sensitivity to variations in eccentricity, for two injection volumes, 3 ml (typical) and 6 ml (extreme). For lenses with no cylindrical power, the sensitivity to variations in eccentricity is most significant. Even in this extreme case, the diameters (a or b) may vary by as much as 300 um and still remain within a 1/8 diopter variation. This sensitivity decreases significantly with the increase in nominal cylindrical power.

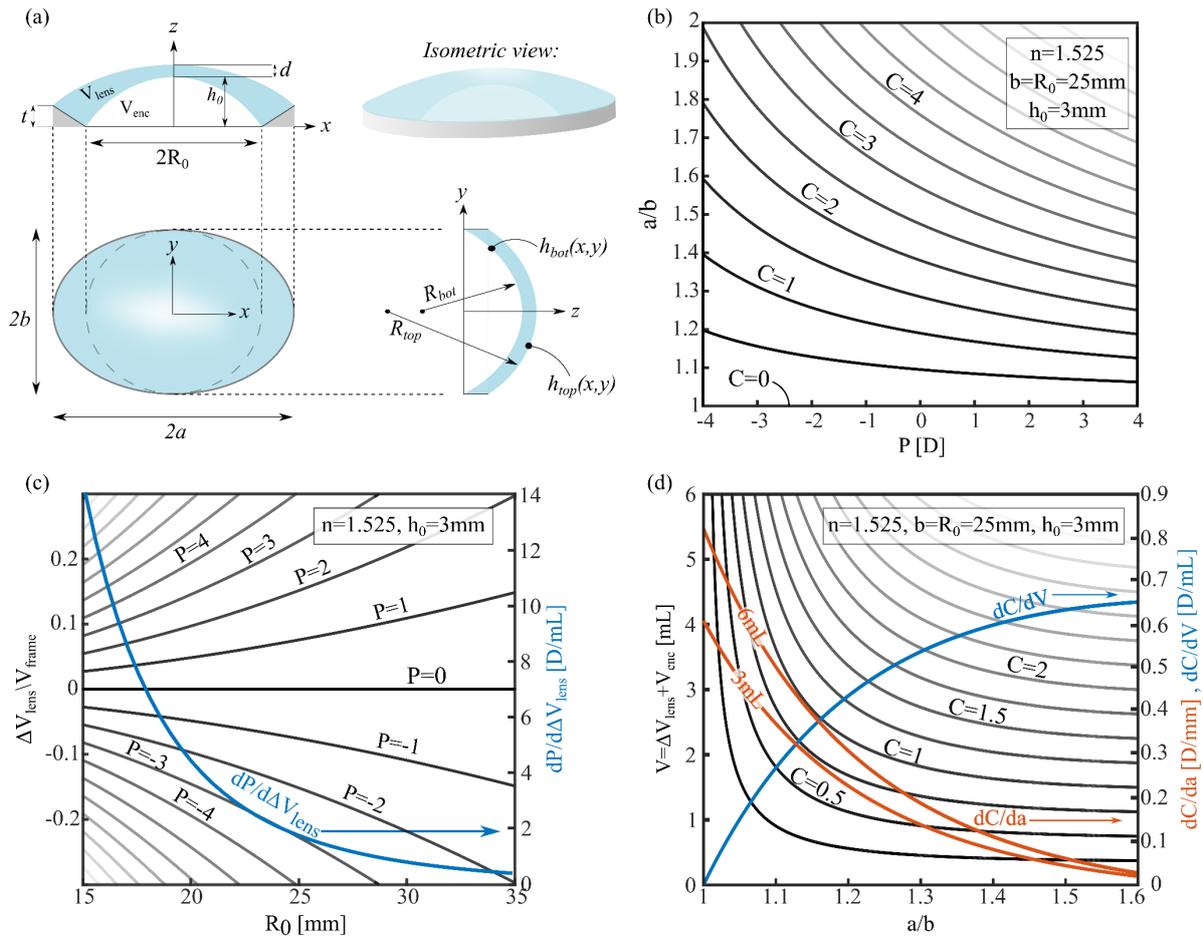

*Figure 3 – Analytical results of the optical powers of the fluidic lens as a function of the frame geometry and the injected volumes.* (a) The examined configuration. The bottom left image shows a top view of the fluidic lens, indicating the semi diameters $a, b$ of the elliptical top contour. The dashed line represents the circular contour beneath it. A cross section along the x axis shows the lens edge thickness $t$, center thickness $d$, the radius of the circular frame $R_0$, the volume of the injected polymer $V_{lens}$ and the volume of the enclosed immersion liquid $V_{enc}$. A cross section along the y axis shows $h_{top}, h_{bot}$ - the shape of the fluidic interfaces, and the radii of curvature of these surfaces $R_{top}, R_{bot}$. (b) The cylindrical diopter of the lens as a function of its spherical diopter and the ratio $a/b$ for $R_0 = 25mm, n = 1.525$. (c) The spherical diopter as a function of the frame radius and the injected polymer liquid, normalized to the frame volume $\Delta V_{lens} / V_{frmae}$ for $n = 1.525$. The blue curve represents the sensitivity of the spherical power to changes in the injected volume. (d) The cylindrical power of the lens as a function of the ratio $a/b$ and the total volume $\Delta V_{lens} + \Delta V_{imm}$, for $b = R_0 = 25mm, n = 1.525$. The blue curve represents the sensitivity of the cylindrical power to changes in the total volume $\Delta V_{lens} + V_{enc}$, and the orange curves represent the sensitivity of the cylindrical power to changes in eccentricity, for total volumes of $3 mL$ (typical) and $6 mL$ (extreme).

**Results**

Figure 4a presents the normalized diopter, $\tilde{P} = P \cdot (\pi R_0^4)/(4(n-1))$, as a function of the injected polymer volume, for 92 lenses fabricated using circular frames, spanning both negative and positive diopters, for two frame diameters, and for two different polymers – one UV curable, and one thermally cured. The error bars in the horizontal direction represent an uncertainty in $\Delta V_{lens}$, which propagates from uncertainties in the injection volume and in the frame dimensions, as detailed in SI5. The power value represents the average power over the entire area of the lens, measured by a Moiré deflectometer (Mapper, Rotlex, Israel). The results are in very good agreement with the theoretical prediction (Equation 3 where $a = b = R_0$, represented by the solid gray line), and demonstrate the ability to design and fabricate lenses covering a wide range of optical powers (here between P=-6 and P=5 diopters. SI Table S1 presents a table of the measured values for all fabricated lenses).

For a given fabricated lens, the quality is determined by the uniformity of optical power over its area. Figures 4b-c present the deviation of the measured spherical and cylindrical power distributions from their expected theoretical values, for two good-quality lenses produced using Fluidic Shaping. The images in Figure 4b correspond to a lens produced in a circular frame $(e = 0, R_0 = 25\,mm)$, i.e., intended to have a purely spherical correction. The resulting mean spherical power in this case deviates by 0.11 diopters (D) from the intended value and the spatial standard deviation is 0.04 D. The lens shows a residual cylindrical power of 0.17 D, with a standard deviation of 0.1D. The cylindrical deviation is within the common acceptable spec, based on the human eye sensitivity[43], and the spherical deviation meets an even more stringent spec of < 1/8 D. The images in Figure 4c correspond to a lens produced in an elliptical frame $(e = 0.552, a = 30\,mm, b = R_0 = 25\,mm)$, with an injected liquid volume designed to obtain the same spherical diopter as in Figure 4b, but with a distinct cylindrical diopter of 1.7 D. This illustrates the ability to independently define the two nominal powers in accordance with Equation 3 and SI4. Here, too, the standard deviation for both powers is within the commonly acceptable ranges. Figure S4 in the SI presents a histogram of power variation over the area of all produced lenses. More than 50% of the lenses show a diopter variation of less than 0.25 D over the entire lens. Another 30% of the lenses show variations of up to 0.4 diopters, and the remaining show larger variations. To improve the yield, the system must be further engineered to guarantee a repeatable process.

An additional important criterion by which optical components are assessed is their surface roughness - lenses with poor surface quality cannot be functionally used. Figure 4d presents an AFM measurement showing an average surface roughness of 1.4 nm in a measurement area of 3μm x 3μm. We have performed such measurements at 9 different locations, on 4 different lenses. As expected, since surface roughness is dictated by surface tension and the molecular structure of the polymer, the result is independent of the location or the shape of the lens. In a total of 9 measurement sites, we obtained RMS values ranging between 0.43 nm and 3.3 nm, with an average value of 1.4 nm (see SI7). These values are between one and two orders of magnitude better than the industry standard which is tens of nanometers[46].

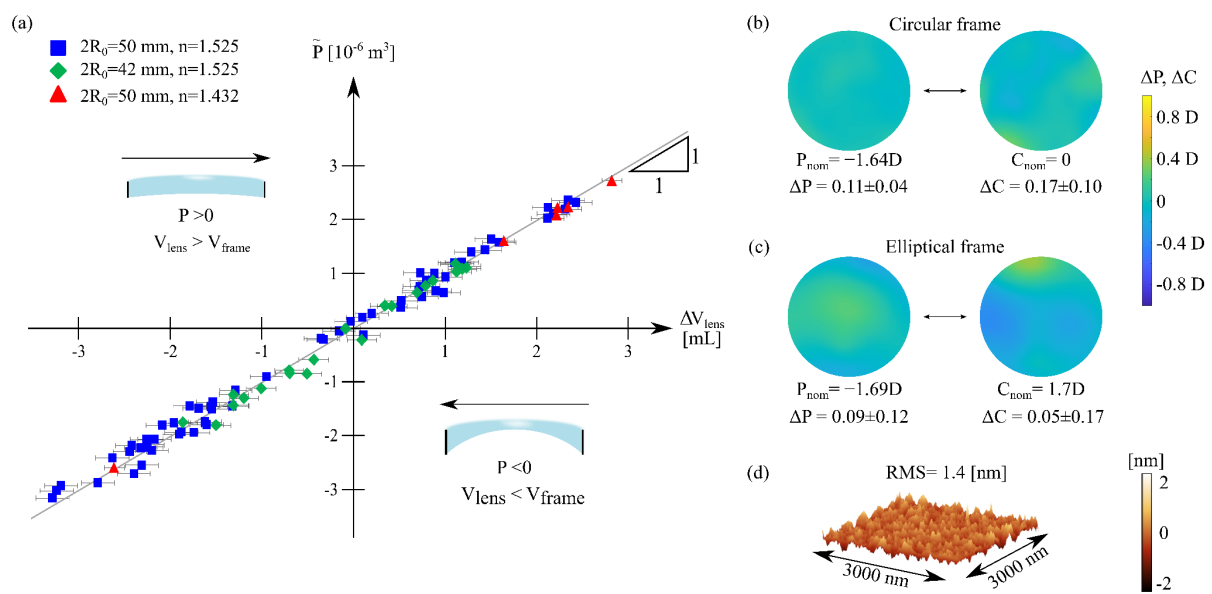

*Figure 4 – Experimental results and characterization of eyewear lenses fabricated using the Fluidic Shaping method.* (a) The normalized measured diopter $\left(\tilde{P} = \frac{\pi R_0^4}{4(n-1)} \cdot P\right)$ as a function of the injected volume $\Delta V_{lens}$, for 92 lenses with three different combinations of frame diameters and refractive indices. The gray line represents the theoretical prediction, showing good agreement with the experimental results. (b) The deviation of the measured spherical and cylindrical diopter maps $P(x,y), C(x,y)$ from their corresponding theoretical values $P_{theory}(x,y), C_{theory}(x,y)$ for a lens fabricated using a circular frame designed to yield a spherical power and no cylindrical power. (c) The power deviations for a lens that was fabricated with an elliptical frame designed to have the same spherical power, but also cylindrical power. The results illustrate the degrees of freedom of the method, and the ability to reach the acceptable power deviations for corrective eyewear. (d) A representative AFM measurement of the surface of the lens, showing nanometric surface roughness (1.4 nm RMS).

**Conclusions**

The lack of access to corrective eyewear in developing countries is an important and long-standing problem. Despite significant commercial and philanthropic efforts, in which pre-fabricated eyewear is imported and distributed, the impact on local communities remains very limited, covering 0.3% of the need. To the best of our knowledge, solving the problem by establishing local manufacturing has been considered non-viable, as the required resources for producing high-quality optics with today's technology are beyond those available in these regions. The method we presented here, which is based on shaping liquid polymer volumes in a small self-contained device, with low energy consumption and minimal waste, offers an entirely new route for the fabrication of prescription lenses. We believe that it can make local manufacturing in low resource settings a viable option and contribute both to global healthcare and the economy.

While primarily aimed at addressing the needs of developing countries, the Fluidic Shaping method also offers an alternative to traditional spectacle lens production methods employed in the developed world. By eliminating the need for any machining or polishing, and by significantly reducing material waste, it well aligns with environmental sustainability goals. The method is also well-position of production at scale, owing to the simplicity of the process and to the short fabrication time per lens.

The surface quality obtained by Fluidic Shaping is consistently outstanding (nanometric) and no further improvement is required to meet the strictest standards. However, to reach power uniformities that compete with those of high-end machining, the yield and automation of the system must be improved. All lenses presented in this paper were manufactured by manual injection, which suffers from inaccuracies in the injected volume and in the pinning of the contact line to the surrounding frame. We hypothesize that automated injection would result in far more consistent results. Moreover, the photopolymer we used is a generic off-the-shelf product, not specifically designed for optics, which undergoes significant (approximately 4%) volumetric shrinkage during solidification. Any non-uniformity in this process leads to local stresses which manifest in deformation of the surface. Thus, polymers with lower shrinkage are expected to improve the lens' quality. While various combinations of polymers and immersion liquids can be considered, materials that are immiscible in water are particularly convenient for Fluidic Shaping as an immersion liquid can be readily matched.

While there is no material loss in the lens fabrication process itself, in our current implementation, the frames are either circular or elliptical and the fitting to eyewear rims still requires 'edging' (i.e. cutting the lens to the required shape), which manifests in a small material loss. In previous work we showed that variations in height along the frame's circumference can be used to create complex freeform optical surfaces[2]. We thus believe that the need for edging can also be eliminated by injecting the polymer to a non-circular frame that matches the eyewear rim and has a varying height such that the desired spherical or cylindrical prescription is maintained. Moreover, the same approach can potentially be used to create progressive lenses with non-uniform optical powers over the lens' area, with deviation from neutral buoyancy serving as an additional degree of freedom[40].

**Acknowledgments.** Funded by the European Union (ERC, Fluidic Shaping, 101044516). Views and opinions expressed are however those of the author(s) only and do not necessarily reflect those of the European Union or the European Research Council Executive Agency. Neither the European Union nor the granting authority can be held responsible for them.

We thank Alexey Razin and Khaled Gommed for their assistance with the experimental setup. We also express our gratitude to Zohar Katzman and Zohar Kadmon from Shamir Optical Industry Ltd. for providing access to their facilities to test our lenses and for their valuable consultation on the results. We thank the '4 Eyes' optometry store in Haifa for mounting our lenses on their eyewear rims.

**Author contributions.** M.E., O.L., V.F., and M.B. conceived the experiments. M.E. developed the theory. M.E. and O.L. designed the experimental setup. M.E. and Y.K. performed the experiments. M.E., O.L., and Y.K. planned the characterization setup. D.W. contributed to sample preparation and assisted with material handling. O.L. and D.W. characterized the material properties. M.E. performed the data analysis. All authors regularly discussed the results and directed the experiments and data analysis. M.E., V.F., and M.B. wrote the manuscript. O.L. reviewed, commented, and corrected the manuscript.

**Competing interests.** The authors declare no competing interests.

**Supplementary information.** Supplementary Information is available for this paper.

**Correspondence and requests for materials** should be addressed to Bercovici M.

# Supplementary Information for

# Fluidic Approach to Corrective Eyewear Manufacturing in Low-Resource Settings

Mor Elgarisi[1], Omer Luria[1], Yotam Katzman[1], Daniel Widerker[1], Valeri Frumkin[1,2], Moran Bercovici[1,*]

[1]Faculty of Mechanical Engineering, Technion – Israel Institute of Technology, Haifa, Israel.

[2]Current affiliation: Department of Mechanical Engineering, Boston University, Boston, MA, USA.

[*]Corresponding author: mberco@technion.ac.il

**This file includes:**

Supplementary Information S1 to S7.

# Table of contents

# SI 1. Description of the experimental setup

The experimental setup shown in Figure S1 consists of a water container that is separated into two parts by a divider plate with a round opening at its center. Both the container and the divider plate are made of PMMA. Resting atop the divider plate is a 3D printed frame (Form Clear, Formlabs, USA), which serves as the boundary for the polymer liquid injection. The frame is secured to the dividing plate on one side with a silicone rubber ring which is compressed by a plastic flange (see 'Zoom A' view in Figure S1). This compression ensures a hermetic seal between the container's upper and lower compartments.

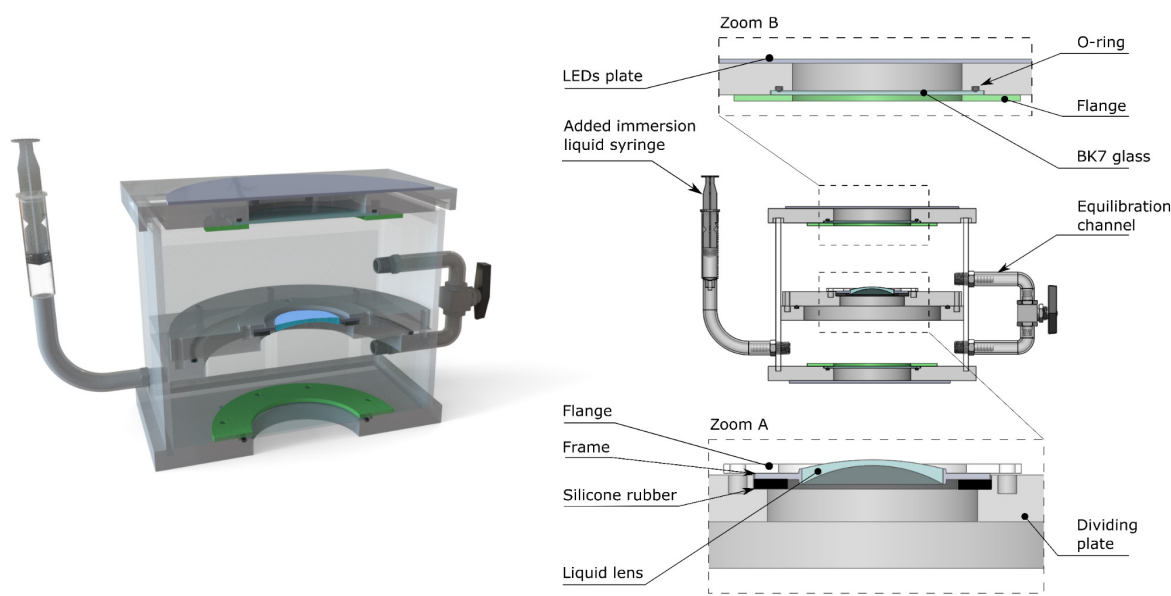

*Figure S1 – **The experimental setup.** Left: a 3D CAD model of the setup (cut at middle plane). Right: A cross-section of the setup and zoom-in views showing its main components.*

In the upper part of the container ('Zoom B' view in Figure S1), there is an opening covered by BK7 glass that allows transmission of UV-A light. The glass is pressed by a flange against an O-ring embedded in the PMMA top cover. Positioned directly above the glass-covered aperture is an LEDs plate, comprising an array of 12 LEDs (SST-10-UV, Mouser, USA). The LEDs emit at a wavelength of 365 nm with an output radiant power of 5 W. A sheet of parchment paper, placed between the flange and the glass (not shown in the figure), acts as an optical diffuser and enables a more uniform illumination over the lens. An identical glass opening and LEDs array are present at the container's lower section, enabling illumination of the lens from both sides.

On one side of the container (right side in Figure S1), there are two openings, one in the upper part and one in the lower part, connected by an "equilibration channel". This channel is open during lens injection and thus equalizes the pressure between the two parts of the container and is closed during lens inflation to allow pressurization of the lower part. On the other side of the container, an additional opening in the lower section connects to a syringe filled with immersion liquid. This syringe is utilized to add an immersion liquid to the bottom part and thus inflate the lens while it's in its liquid state.

Polymer liquids and immersion liquid:

We used two different materials as the lens liquid - a photocurable polymer (UV resin VidaRosa J-2D-UVDJ250G, Dongguan, China) with a refractive index of 1.52 and a specific density of 1.07, and polydimethylsiloxane (PDMS, Sylgard 184, Dow, USA) with a refractive index of 1.42 and a specific density of 1.04. The immersion liquid for both materials was a mixture of DI water with glycerol where for the VidaRosa we used 27% glycerol dissolved in water and for the PDMS we used 11%.

## SI 2. Derivation of the liquid interfaces' shapes

Figure S2 presents a schematic illustration of a fluidic lens formed within a frame of height $t$, consisting of a circular bottom contour of radius $R_0$ and an elliptical top contour with semi diameters $a, b$. The volume of the liquid photopolymer injected into the frame is $V_{lens}$, and the volume enclosed between its bottom surface and the x-y plane is $V_{enc}$.

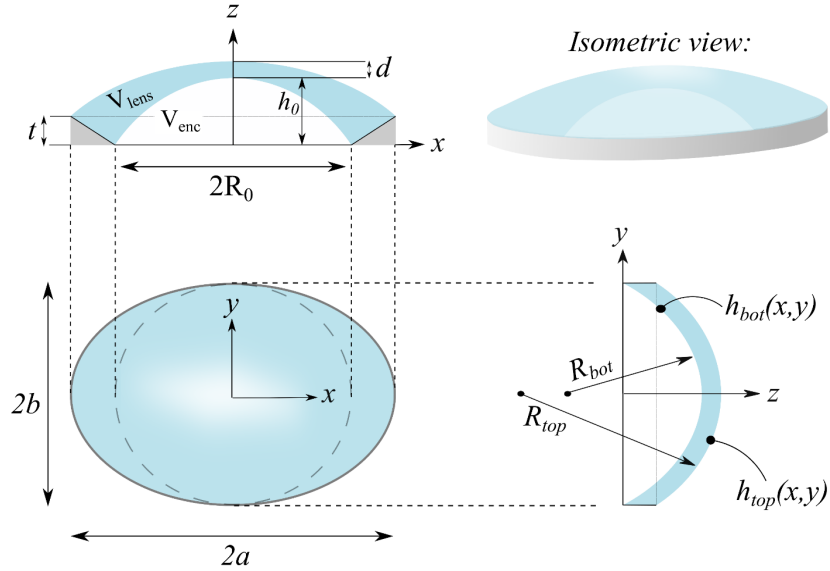

*Figure S2 - Schematic illustration of the examined configuration.*

Under neutral buoyancy conditions, where the density of the immersion liquid that surrounds the optical polymer liquid is equal to that of the polymer, changes in the shape of the interfaces affect the energy of the system only through surface tension.

At steady state, the fluidic interfaces will take a shape that minimizes the free energy of the system:

$$\Pi = \iint_\Gamma F(x,y)dxdy$$

$$F = \gamma\sqrt{1+\left(h_x^{(t)}\right)^2+\left(h_y^{(t)}\right)^2} + \gamma\sqrt{1+\left(h_x^{(b)}\right)^2+\left(h_y^{(b)}\right)^2} + \lambda^{(t)}\left(h^{(t)}-t\right)+\lambda^{(b)}h^{(b)} \quad (1)$$

,

where the superscripts (t) and (b) refer to the top and bottom surfaces, respectively, and $\lambda^{(t)}, \lambda^{(b)}$ are the Lagrange multipliers associated with the known volume constraints:

$$V_{lens} + V_{enc} - V_{frame} = \iint_\Gamma \left( h^{(t)} - t \right) dxdy \ ; \ V_{enc} = \iint_\Gamma h^{(b)} dxdy , \qquad (2)$$

where $V_{frame}$ is the volume encompassed within the frame: $V_{frame} = \frac{1}{2}\pi t(a+b) R_0$.

Minimization of the free energy requires the first variation of the energy potential to vanish, which yields the standard Euler-Lagrange equations[1]

$$\begin{cases} \frac{\partial F}{\partial h^{(t)}} - \frac{d}{dx}\frac{\partial F}{\partial h_x^{(t)}} - \frac{d}{dy}\frac{\partial F}{\partial h_y^{(t)}} = 0 \\ \frac{\partial F}{\partial h^{(b)}} - \frac{d}{dx}\frac{\partial F}{\partial h_x^{(b)}} - \frac{d}{dy}\frac{\partial F}{\partial h_y^{(b)}} = 0 \end{cases} \Rightarrow \begin{cases} \lambda^{(t)} + \gamma \dfrac{2h_{xy}^{(t)} h_x^{(t)} h_y^{(t)} - h_{xx}^{(t)}\left(1+\left(h_y^{(t)}\right)^2\right) - h_{yy}^{(t)}\left(1+\left(h_x^{(t)}\right)^2\right)}{\left(1+\left(h_x^{(t)}\right)^2+\left(h_y^{(t)}\right)^2\right)^{3/2}} = 0 \\ \lambda^{(b)} + \gamma \dfrac{2h_{xy}^{(b)} h_x^{(b)} h_y^{(b)} - h_{xx}^{(b)}\left(1+\left(h_y^{(b)}\right)^2\right) - h_{yy}^{(b)}\left(1+\left(h_x^{(b)}\right)^2\right)}{\left(1+\left(h_x^{(b)}\right)^2+\left(h_y^{(b)}\right)^2\right)^{3/2}} = 0 \end{cases}. \qquad (3)$$

We define the following dimensionless variables: $X = \frac{x}{a}, Y = \frac{y}{b}, H^{(t)} = \frac{h^{(t)}}{d_c}, H^{(b)} = \frac{h^{(b)}}{d_c}$. Since the equations for both surfaces are identical, we here drop the (b) and (t) superscript notation, and express equation (3) for both the top and bottom surfaces as

$$\frac{\lambda a^2}{d_c} + \gamma \frac{2\dfrac{d_c^2}{b^2} H_{XY} H_X H_Y - H_{XX}\left(1+\dfrac{d_c^2}{b^2}H_Y^2\right) - \dfrac{a^2}{b^2}H_{YY}\left(1+\dfrac{d_c^2}{a^2}H_X^2\right)}{\left(1+\dfrac{d_c^2}{a^2}H_X^2+\dfrac{d_c^2}{b^2}H_Y^2\right)^{3/2}} = 0. \qquad (4)$$

Since we are interested in optical components, for which the characteristic heights are an order of magnitude lower than the component length, i.e. $\frac{d_c^2}{a^2}, \frac{d_c^2}{b^2} \ll 1$, equation (4) can be linearized to yield

$$H_{XX} + \frac{a^2}{b^2} H_{YY} = \frac{\lambda a^2}{\gamma d_c}, \qquad (5)$$

and in dimensional form for both surfaces

$$\begin{aligned} h_{xx}^{(t)} + h_{yy}^{(t)} &= K^{(t)} \\ h_{xx}^{(b)} + h_{yy}^{(b)} &= K^{(b)} \end{aligned}, \qquad (6)$$

where $K^{(t)}$ and $K^{(b)}$ are unknown constants resulting from the Lagrange multipliers.

Bottom surface:

The boundary of the bottom surface is circular. Hence, it is convenient to write the equation and the boundary conditions in a radial coordinate system whose origin coincides with the cartesian one. Under the assumption of axisymmetry, the equation takes the form

$$\frac{1}{r}h_r^{(b)} + h_{rr}^{(b)} = K^{(b)}$$
$$\begin{cases} h^{(b)}(r=R_0) = 0 \\ h_r^{(b)}(r=0) = 0 \end{cases}, \qquad (7)$$

for which the solution is

$$h^{(b)} = \frac{1}{4}K^{(b)}\left(r^2 - R_0^2\right). \qquad (8)$$

Using the volume constraint, $V_{enc} = 2\pi \int_0^{R_0} h^{(b)} r dr$, the constant $K^{(b)}$ can be found, resulting in a closed form expression for the bottom liquid interface,

$$h^{(b)} = \frac{2V_{enc}}{\pi R_0^2}\left(1 - \frac{r^2}{R_0^2}\right). \qquad (9)$$

Top surface:

The top surface, in its most general form, consists of an elliptical boundary. To find the top surface analytically, we transform equation (6) to an elliptical coordinate system, by defining $x = c \cdot \cosh(\mu)\cos(\upsilon)$, $y = c \cdot \sinh(\mu)\sin(\upsilon)$, where $c = \sqrt{a^2 - b^2}$ is the distance of the ellipse foci from the origin. For each value of $\mu$, $x(\upsilon)$, $y(\upsilon)$ represent a confocal ellipse, which for $\mu_0 = \text{arccosh}\left(\frac{a}{c}\right)$ satisfies the boundary $\frac{x^2}{a^2} + \frac{y^2}{b^2} = 1$. With the elliptical coordinate system, equation (6) takes the form[2]

$$h_{\mu\mu}^{(t)} + h_{\upsilon\upsilon}^{(t)} = K\left(\cosh^2\mu - \cos^2\upsilon\right) = \frac{1}{2}K\left(\cosh(2\mu) - \cos(2\upsilon)\right), \qquad (10)$$

where $K = K^{(t)}c^2$. We solve the equation by separating it into particular and homogenous solutions.

The particular solution is

$$h_p^{(t)} = \frac{1}{8}K\left(\cosh(2\mu) + \cos(2\upsilon)\right). \tag{11}$$

For the homogenous solution, we define a separation of variables with $h_h^{(t)} = M(\mu)N(\upsilon)$, and solve

$$NM_{\mu\mu} + MN_{\upsilon\upsilon} = 0. \tag{12}$$

Combining the particular and homogenous solutions yields

$$h^{(t)} = \sum_0^\infty \left[\left(A_n \cos(n\upsilon) + B_n \sin(n\upsilon)\right)\left(C_n \cosh(n\mu) + D_n \sinh(n\mu)\right)\right] + \\ + \frac{K}{8}\left(\cosh(2\mu) + \cos(2\upsilon)\right). \tag{13}$$

In order to satisfy the continuity of the top surface $h^{(t)}$ and its derivative at $\mu = 0$: $h(0,\upsilon) = h(0,-\upsilon)$, $\frac{\partial h}{\partial \mu}(0,\upsilon) = -\frac{\partial h}{\partial \mu}(0,-\upsilon)$, equation (15) reduces to

$$h^{(t)} = \sum_0^\infty \left[\left(A_n^* \cosh(n\mu)\cos(n\upsilon) + B_n^* \sin(n\upsilon)\sinh(n\mu)\right)\right] + \frac{K}{8}\left(\cosh(2\mu) + \cos(2\upsilon)\right). \tag{14}$$

To satisfy the boundary condition, $h^{(t)}(\mu = \mu_0, \upsilon) = t$: $A_0^* = t - \frac{K}{8}\cosh(2\mu_0)$, $A_2^* = -\frac{K}{8\cosh(2\mu_0)}$, and all other coefficients must vanish,

$$h^{(t)} = \frac{K}{8}\left(\cosh(2\mu) - \cosh(2\mu_0)\right)\left(1 - \frac{\cos(2\upsilon)}{\cosh(2\mu_0)}\right) + t. \tag{15}$$

The constant $K$ can be calculated using the constraint of the total volume,

$$V_{lens} + V_{enc} - V_{frame} = \iint_\Gamma \left(h^{(top)} - t\right)dxdy = \int_0^{2\pi}\int_0^{\mu_0}\left(h(\mu,\upsilon) - t\right)\left(\sinh^2\mu + \sin^2\upsilon\right)c^2 d\mu d\upsilon, \tag{16}$$

which yields

$$K = \frac{32\left(V_{lens} + V_{enc} - V_{frame}\right)}{\pi c^2 \left(\tanh(2\mu_0) - \cosh(2\mu_0)\sinh(2\mu_0)\right)}. \tag{17}$$

Using the relations $\cosh(2\mu_0) = \frac{2ab}{c^2}$, $\sinh(2\mu_0) = \frac{a^2 + b^2}{c^2}$ and defining $\Delta V_{lens} = V_{lens} - V_{frame}$, Equation (17) simplifies to

$$K = -\frac{4\left(a^4 - b^4\right)\left(\Delta V_{lens} + V_{enc}\right)}{\pi a^3 b^3}. \tag{18}$$

Using the transformation relation from elliptical coordinates to cartesian coordinates[3], the solution for the top liquid interface is

$$h^{(t)} = \frac{\left(\Delta V_{lens} + V_{enc}\right)}{2\pi a^3 b^3}\left(x^2 + y^2 - \left(a^2 + b^2\right) + \sqrt{\left(x^2 + y^2\right)^2 + c^4 - 2c^2\left(x^2 - y^2\right)}\right) \times$$

$$\times \left(\frac{4c^2 x^2 - 2a^2\left(x^2 + y^2 + c^2 + \sqrt{\left(x^2 + y^2\right)^2 + c^4 - 2c^2\left(x^2 - y^2\right)}\right)}{x^2 + y^2 + c^2 + \sqrt{\left(x^2 + y^2\right)^2 + c^4 - 2c^2\left(x^2 - y^2\right)}}\right) + t. \tag{19}$$

For the axisymmetric case of $c = 0$, i.e., a circular top contour with $a = b = R_0$, the height of the top surface will take the shape

$$h^{(t)}\Big|_{c \to 0} = \frac{2\left(\Delta V_{lens} + V_{enc}\right)}{\pi R_0^2}\left(1 - \frac{r^2}{R_0^2}\right) + t. \tag{20}$$

## SI 3. Calculating the optical powers of the lens

The Fluidic Shaping method allows to design a specific lens diopter, simply by controlling the volume of the injected polymer $V_{lens}$ and the volume of the added immersion liquid $V_{enc}$, for a given set of frame dimensions $a, b, R_0, t$ and the refractive index of the liquid polymer, $n$. In this section, we derive explicit expressions for the spherical and cylindrical powers of the lens as a function of the known parameters.

The spherical diopter $P$ and the cylindrical diopter $C$ of a lens can be expressed using the 'Lens Maker' equations[4]:

$$P = (n-1)\left(\frac{1}{\bar{R}_{top}} - \frac{1}{\bar{R}_{bot}} + \frac{(n-1)d}{n\bar{R}_{top}\bar{R}_{bot}}\right)$$
$$C = (n-1)\left(\frac{1}{R_{xtop}} - \frac{1}{R_{ytop}}\right)\left(1 + \frac{(n-1)d}{n\bar{R}_{bot}}\right), \qquad (21)$$

where $\bar{R}_{bot}$ and $\bar{R}_{top}$ are the mean radii of curvatures of the bottom and top surfaces respectively, $R_{xtop}$ and $R_{ytop}$ are the radii of curvature of the top surface along the x and y axes, and $d$ is the center thickness of the lens.

Using the solutions for the liquid interface shapes from Equations (9) and (19), the radii of curvatures and the lens thickness can be calculated:

$$\frac{1}{R_{x,top}} = \left|\frac{h_{xx}^{(t)}}{\left(1+\left(h_x^{(t)}\right)^2\right)^{3/2}}\right| = \frac{4\left(V_{lens}+V_{enc}-V_{frame}\right)}{\pi a^3 b}$$

$$\frac{1}{R_{y,top}} = \left|\frac{h_{yy}^{(t)}}{\left(1+\left(h_y^{(t)}\right)^2\right)^{3/2}}\right| = \frac{4\left(V_{lens}+V_{enc}-V_{frame}\right)}{\pi a b^3}$$

$$\frac{1}{\bar{R}_{top}} = \frac{1}{2}\left(\frac{1}{R_{x,top}} + \frac{1}{R_{y,top}}\right) = \frac{2\left(V_{lens}+V_{enc}-V_{frame}\right)}{\pi a^3 b^3}\left(a^2+b^2\right) \qquad (22)$$

$$\frac{1}{\bar{R}_{bot}} = \left|\frac{h_{rr}^{(b)}}{\left(1+\left(h_r^{(b)}\right)^2\right)^{3/2}}\right| = \frac{4V_{enc}}{\pi R_0^4}$$

$$d = h^{(t)}(0,0) - h^{(b)}(0) = t + \frac{2}{\pi}\left(\frac{\left(V_{lens}+V_{enc}-V_{frame}\right)}{ab} - \frac{V_{enc}}{R_0^2}\right).$$

Using equations (21) and (22), the spherical power of the lens can be expressed as

$$P = 2(n-1)\left[\frac{(\Delta V_{lens} + V_{enc})(a^2 + b^2)}{\pi a^3 b^3} - \frac{2V_{enc}}{\pi R_0^4} + \right.$$
$$\left. + 4\frac{(n-1)}{n}\frac{(\Delta V_{lens} + V_{enc})(a^2 + b^2)V_{enc}}{\pi^2 a^3 b^3 R_0^4}\left(t + \frac{2}{\pi}\left(\frac{(\Delta V_{lens} + V_{enc})}{ab} - \frac{V_{enc}}{R_0^2}\right)\right)\right],$$
(23)

and similarly, the cylindrical power can be expressed as

$$C = (n-1)\frac{4(\Delta V_{lens} + V_{enc})(a^2 - b^2)}{\pi a^3 b^3}\left(1 + \frac{(n-1)}{n}\frac{4V_{enc}}{\pi R_0^4}\left(t + \frac{2}{\pi}\left(\frac{(\Delta V_{lens} + V_{enc})}{ab} - \frac{V_{enc}}{R_0^2}\right)\right)\right). \quad (24)$$

Under the assumption of thin lenses, acceptable for eyewear lenses with a typical radius of tens of millimeters and a center thickness of a few millimeters[4] (which holds well for the lenses we fabricate with our method), a simplified expression for the spherical and cylindrical powers can be obtained,

$$(a) \quad P = 2(n-1)\left(\frac{(\Delta V_{lens} + V_{enc})(a^2 + b^2)}{\pi a^3 b^3} - \frac{2V_{enc}}{\pi R_0^4}\right)$$
$$(b) \quad C = 4(n-1)\frac{(\Delta V_{lens} + V_{enc})(a^2 - b^2)}{\pi a^3 b^3}$$
(25)

For the case where the frame is circular on both sides, i.e., $a = b = R_0$, there will be no cylindrical power, and the spherical power will be expressed as

$$P = \frac{4(n-1)}{\pi R_0^4}\Delta V_{lens}. \quad (26)$$

# SI 4. A flowchart for calculating the frame geometry and liquid volumes needed for fabricating a desired lens

The Fluidic Shaping method enables the fabrication of eyewear lenses with any combination of spherical and cylindrical powers. The following flowchart outlines the processes for obtaining the required frame geometry and liquid volumes needed to fabricate lenses with specific optical powers (see details below the flowchart).

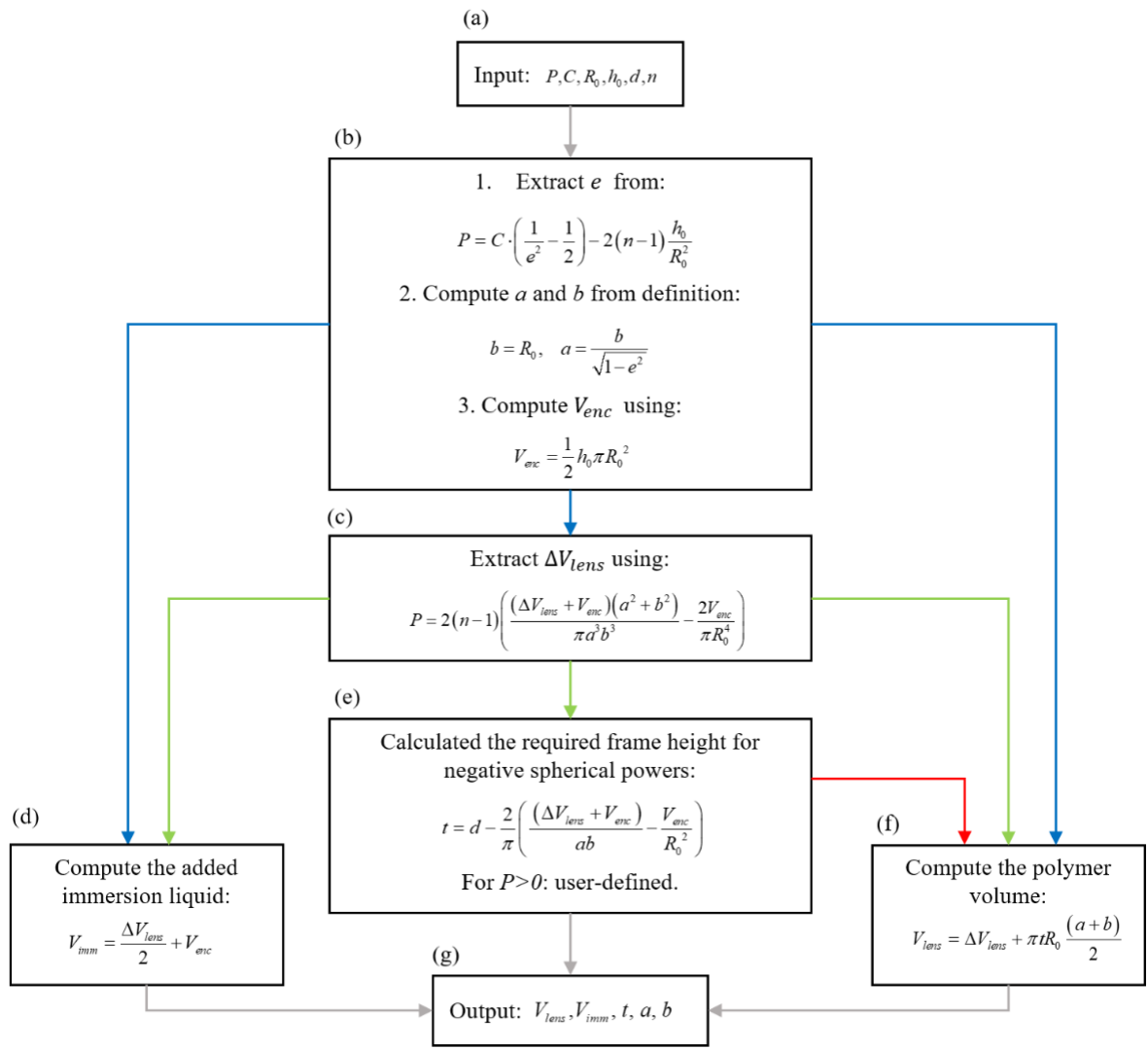

*Figure S3 – Flowchart for calculating the frame geometry and liquid volumes for fabricating a desired lens.*

(a) The required input parameters are the spherical power ($P$), cylindrical power ($C$), desired size of the lens ($R_0$), the clearance from the eye ($h_0$), the refractive index ($n$) of the polymer liquid, and for negative lenses- the allowed lens thickness (*d*). For positive lenses: the frame thickness (*t*).

(b) Using the input parameters, calculate the frame's eccentricity using the term $P = C \cdot \left( \frac{1}{e^2} - \frac{1}{2} \right) - 2(n-1)\frac{h_0}{R_0^2}$, which results from combining Equations (25a) and (25b). This calculation yields the elliptical boundary semi diameters (*a, b*). In addition, calculate the volume enclosed under the lens' bottom surface, $V_{enc}$, using Equation (9).

(c) Calculate $\Delta V_{lens}$ using Equation (25a).

(d) Using the enclosed immersion liquid volume $(V_{enc})$, and the value of $\Delta V_{lens}$, calculate the total volume of immersion liquid that should be injected, $V_{imm} = \frac{1}{2}\Delta V_{lens} + V_{enc}$.

(e) For negative spherical power lenses, calculate the height of the bounding frame using Equation (22). For positive spherical power, this value is one of the inputs.

(f) Using the frame geometry, $\Delta V_{lens}$, and the enclosed immersion liquid volume, $V_{enc}$, calculate the volume of polymer liquid that should be injected, $V_{lens}$.

(g) The output is then the frame geometry $a, b, t$ and the required volumes $V_{lens}, V_{imm}$.

## SI 5. Uncertainties in the injection volume and in the frame dimensions

Figure 4a in the manuscript shows the measurements of the spherical power of the lenses that were fabricated in our experimental setup, as a function of $\Delta V_{lens} = V_{lens} - V_{frame}$. The error bars shown in this figure represent an uncertainty in $\Delta V_{lens}$, which propagates from uncertainties in the injection volume and in the frame dimensions $(R_0, t)$. We express $\Delta V_{lens}$ as

$$\Delta V_{lens} = V_{lens} - V_{frame} = \frac{m}{\rho} - \pi R_0^2 t, \qquad (27)$$

where $m$ is the mass of the lens, $\rho$ is its (liquid) density, $R_0$ is the inner radius of the frame, and $t$ is its height. The error in $\Delta V_{lens}$ can thus be defined from error propagation as

$$\delta(\Delta V_{lens}) = \sqrt{\left(\frac{\partial(\Delta V_{lens})}{\partial m}\delta m\right)^2 + \left(\frac{\partial(\Delta V_{lens})}{\partial \rho}\delta \rho\right)^2 + \left(\frac{\partial(\Delta V_{lens})}{\partial R_0}\delta R_0\right)^2 + \left(\frac{\partial(\Delta V_{lens})}{\partial t}\delta t\right)^2}, \qquad (28)$$

which by can be simplified using derivatives of $\Delta V_{lens}$ to the form

$$\delta V = \sqrt{\left(\frac{\delta m}{\rho}\right)^2 + \left(\frac{m\delta \rho}{\rho^2}\right)^2 + \left(2\pi R_0 t \delta R_0\right)^2 + \left(\pi R_0^2 \delta t\right)^2}. \qquad (29)$$

$\delta m$ represents the variation in the mass of the lens. To obtain an estimate for this value, we weigh each lens (still on its frame) after solidification, and subtract the weight of the empty frame. Since the error of the scale is on the order of 0.01 gr, the most significant contribution to error is the weight of the empty frame – the largest weight difference between two frames of the same type was 0.15 gr (measured over 20 frames). The density $\rho$ was measured using a standard hydrometer (B61801, SP Bel-Art, USA) with an accuracy of $\delta\rho = 0.001 \ gr/cm^3$. The radius and height of the frame were measured using a caliper with an accuracy of $\delta R_0 = \delta t = 0.05 mm$.

For photocurable polymer with $\rho = 1.07 \ g/cm^3$, and a typical frame height of 5mm, the uncertainty in volume will be 0.18 mL for a 50 mm diameter and 0.16 mL for a 42 mm diameter. For PDMS, with $\rho = 1.04 \ g/cm^3$ and with 0.01 gr uncertainty in mass (the PDMS lens can be detached from its frame and measured separately, thus the error is smaller), the uncertainty in volume for a lens of the same dimensions is 0.11 mL.

## SI 6. Experimental results – additional data

Spherical lenses:

Figure 4a in the manuscript shows the measurements of the spherical diopter of the lenses that were fabricated in our experimental setup. The following table lists all of the measured lenses together with their measured average power and the standard deviation of power over the entire area of the lens.

*Table S1 – Fabricated circular-frame lenses, measured mean and STD of spherical power.*

| # | Diameter $2R_0$ [mm] | Width $t$ [mm] | Polymer | $\Delta V_{lens} = V_{lens} - V_{frame}$ [mL] | Measured power [1/m] | Measured STD [1/m] | Measured normalized power $\left( \tilde{P} = \frac{\pi R_0^4}{4(n-1)} \cdot P \right)$ | Theoretical normalized power $\tilde{P}_{theory}$ |
|---|---|---|---|---|---|---|---|---|
| 1 | 42 | 4.1 | VidaRosa | 1.19 | 3.74 | 0.376 | 1.10 | 1.19 |
| 2 | 42 | 3.9 | VidaRosa | 0.78 | 2.67 | 0.169 | 0.78 | 0.78 |
| 3 | 42 | 6.1 | VidaRosa | -1.50 | -6.12 | 0.246 | -1.80 | -1.50 |
| 4 | 42 | 4 | VidaRosa | 0.42 | 1.38 | 0.084 | 0.40 | 0.42 |
| 5 | 42 | 4.05 | VidaRosa | 1.13 | 3.56 | 0.204 | 1.05 | 1.13 |
| 6 | 42 | 4.15 | VidaRosa | 1.23 | 3.79 | 0.115 | 1.11 | 1.23 |
| 7 | 42 | 4 | VidaRosa | 0.69 | 2.25 | 0.315 | 0.66 | 0.69 |
| 8 | 42 | 4 | VidaRosa | -1.20 | -4.41 | 0.648 | -1.29 | -1.20 |
| 9 | 42 | 4 | VidaRosa | 1.11 | 4.06 | 0.297 | 1.19 | 1.11 |
| 10 | 42 | 3.8 | VidaRosa | -0.09 | -0.03 | 0.320 | -0.01 | -0.09 |
| 11 | 42 | 6.1 | VidaRosa | -0.51 | -2.86 | 0.245 | -0.84 | -0.51 |
| 12 | 42 | 6.1 | VidaRosa | -0.43 | -1.98 | 0.246 | -0.58 | -0.43 |
| 13 | 42 | 3.8 | VidaRosa | -0.69 | -2.83 | 0.803 | -0.83 | -0.69 |
| 14 | 42 | 5.7 | VidaRosa | -1.31 | -4.17 | 0.150 | -1.23 | -1.31 |
| 15 | 42 | 4.1 | VidaRosa | -1.01 | -3.79 | 0.430 | -1.11 | -1.01 |
| 16 | 42 | 6 | VidaRosa | 0.86 | 2.99 | 0.246 | 0.88 | 0.86 |
| 17 | 42 | 3.8 | VidaRosa | 0.34 | 1.42 | 0.111 | 0.42 | 0.34 |
| 18 | 45 | 4.85 | VidaRosa | -1.86 | -4.51 | 0.399 | -1.75 | -1.86 |
| 19 | 45 | 4.95 | VidaRosa | -0.70 | -2.12 | 0.298 | -0.78 | -0.70 |
| 20 | 45 | 4.95 | VidaRosa | -1.31 | -3.89 | 0.271 | -1.43 | -1.31 |
| 21 | 42 | 6.2 | VidaRosa | 0.09 | -0.73 | 0.324 | -0.21 | 0.09 |
| 22 | 50 | 5.5 | VidaRosa | -1.90 | -3.33 | 0.191 | -1.96 | -1.90 |
| 23 | 50 | 5.65 | VidaRosa | -2.31 | -4.29 | 0.245 | -2.53 | -2.31 |
| 24 | 50 | 5.6 | VidaRosa | -1.60 | -3.03 | 0.337 | -1.79 | -1.60 |
| 25 | 50 | 4.05 | VidaRosa | -1.88 | -3.25 | 0.314 | -1.92 | -1.88 |
| 26 | 50 | 5.6 | VidaRosa | -1.56 | -2.45 | 0.328 | -1.45 | -1.56 |
| 27 | 50 | 5.6 | VidaRosa | -1.32 | -2.44 | 0.483 | -1.44 | -1.32 |
| 28 | 50 | 3.9 | VidaRosa | 2.13 | 3.79 | 0.151 | 2.24 | 2.13 |
| 29 | 50 | 3.9 | VidaRosa | 2.12 | 3.45 | 0.281 | 2.04 | 2.12 |
| 30 | 50 | 4 | VidaRosa | 1.18 | 2.08 | 0.202 | 1.23 | 1.18 |
| 31 | 50 | 3.8 | VidaRosa | -1.54 | -2.31 | 0.292 | -1.37 | -1.54 |
| 32 | 50 | 3.8 | VidaRosa | -1.28 | -2.02 | 0.345 | -1.15 | -1.28 |
| 33 | 50 | 3.9 | VidaRosa | 1.29 | 2.40 | 0.131 | 0.89 | 0.80 |
| 34 | 50 | 5.6 | VidaRosa | -2.26 | -3.49 | 0.247 | 1.41 | 1.29 |
| 35 | 50 | 4 | VidaRosa | 0.52 | 0.87 | 0.032 | -2.06 | -2.26 |
| 36 | 50 | 3.95 | VidaRosa | 1.59 | 2.70 | 0.119 | 0.51 | 0.52 |
| 37 | 50 | 3.85 | VidaRosa | 0.73 | 1.74 | 0.228 | 1.59 | 1.59 |
| 38 | 50 | 3.95 | VidaRosa | 2.31 | 3.73 | 0.475 | 1.03 | 0.73 |
| 39 | 50 | 3.9 | VidaRosa | 1.44 | 2.46 | 0.179 | 2.20 | 2.31 |
| 40 | 50 | 3.9 | VidaRosa | 2.34 | 4.02 | 0.374 | 1.45 | 1.44 |
| 41 | 50 | 5.6 | VidaRosa | -1.74 | -3.27 | 0.424 | 2.37 | 2.34 |
| 42 | 50 | 3.9 | VidaRosa | 0.10 | 0.35 | 0.242 | -1.93 | -1.74 |
| 43 | 50 | 5.8 | VidaRosa | 0.71 | 1.19 | 0.184 | 0.21 | 0.10 |
| 44 | 50 | 3.85 | VidaRosa | 1.51 | 2.80 | 0.298 | 0.70 | 0.71 |
| 45 | 50 | 3.8 | VidaRosa | -0.95 | -1.64 | 0.038 | 1.65 | 1.51 |
| 46 | 50 | 5.6 | VidaRosa | -2.26 | -3.73 | 0.144 | -0.89 | -0.95 |
| 47 | 50 | 3.8 | VidaRosa | 0.20 | 0.47 | 0.271 | -2.20 | -2.26 |
| 48 | 50 | 4 | VidaRosa | 0.74 | 1.00 | 0.173 | 0.28 | 0.20 |
| 49 | 50 | 5.5 | VidaRosa | -1.69 | -2.51 | 0.127 | 0.59 | 0.74 |
| 50 | 50 | 5.85 | VidaRosa | 1.13 | 2.03 | 0.208 | -1.48 | -1.69 |
| 51 | 50 | 5.6 | VidaRosa | -2.20 | -3.83 | 0.366 | 1.20 | 1.13 |
| 52 | 50 | 5.6 | VidaRosa | -2.63 | -4.06 | 0.413 | -2.26 | -2.20 |
| 53 | 50 | 3.9 | VidaRosa | 2.18 | 3.59 | 0.361 | -2.40 | -2.63 |
| 54 | 50 | 5.6 | VidaRosa | -1.96 | -2.97 | 0.638 | 2.12 | 2.18 |
| 55 | 50 | 4.1 | VidaRosa | 0.89 | 1.16 | 0.107 | -1.75 | -1.96 |
| 56 | 50 | 4.1 | VidaRosa | 0.99 | 1.11 | 0.073 | 0.69 | 0.89 |
| 57 | 50 | 3.85 | VidaRosa | 1.10 | 2.06 | 0.177 | 0.66 | 0.99 |
| 58 | 50 | 3.8 | VidaRosa | -1.79 | -2.43 | 0.642 | 1.22 | 1.10 |
| 59 | 50 | 5.3 | VidaRosa | -2.17 | -3.48 | 0.430 | -1.43 | -1.79 |
| 60 | 50 | 5.6 | VidaRosa | -2.79 | -4.85 | 0.573 | -2.05 | -2.17 |
| 61 | 50 | 4 | VidaRosa | 1.21 | 1.95 | 0.149 | -2.86 | -2.79 |
| 62 | 50 | 3.95 | VidaRosa | 0.72 | 1.32 | 0.135 | 1.15 | 1.21 |
| 63 | 50 | 4.15 | VidaRosa | 0.52 | 0.65 | 0.104 | 0.78 | 0.72 |
| 64 | 50 | 3.8 | VidaRosa | -0.36 | -0.32 | 0.139 | 0.38 | 0.52 |
| 65 | 50 | 3.95 | VidaRosa | -0.03 | 0.22 | 0.428 | -0.19 | -0.36 |
| 66 | 50 | 3.9 | VidaRosa | 0.88 | 1.72 | 0.097 | 0.13 | -0.03 |
| 67 | 50 | 5.5 | VidaRosa | -1.55 | -2.46 | 0.253 | 1.02 | 0.88 |
| 68 | 50 | 4 | VidaRosa | 1.01 | 1.61 | 0.147 | -1.45 | -1.55 |
| 69 | 50 | 5 | VidaRosa | -1.55 | -2.53 | 0.364 | 0.95 | 1.01 |
| 70 | 50 | 3.9 | VidaRosa | 2.43 | 3.95 | 0.137 | -1.49 | -1.55 |
| 71 | 50 | 4.85 | VidaRosa | -2.42 | -3.68 | 0.209 | 2.33 | 2.43 |
| 72 | 50 | 4.8 | VidaRosa | -1.62 | -2.94 | 0.195 | -2.17 | -2.42 |
| 73 | 50 | 5.8 | VidaRosa | -2.32 | -3.75 | 0.180 | -1.74 | -1.62 |
| 74 | 50 | 5 | VidaRosa | -2.25 | -3.65 | 0.171 | -2.21 | -2.32 |
| 75 | 50 | 5 | VidaRosa | -1.62 | -2.95 | 0.232 | -2.16 | -2.25 |
| 76 | 50 | 5 | VidaRosa | -2.44 | -3.87 | 0.410 | -1.74 | -1.62 |
| 77 | 50 | 5 | VidaRosa | -2.09 | -3.04 | 0.384 | -2.28 | -2.44 |
| 78 | 50 | 6 | VidaRosa | -2.40 | -4.56 | 0.289 | -1.79 | -2.09 |
| 79 | 50 | 5.6 | VidaRosa | -3.24 | -5.10 | 0.468 | -2.69 | -2.40 |
| 80 | 50 | 5.6 | VidaRosa | -3.29 | -5.33 | 0.708 | -3.01 | -3.24 |
| 81 | 50 | 5.6 | VidaRosa | -3.19 | -4.94 | 0.426 | -3.14 | -3.29 |
| 82 | 50 | 3.85 | VidaRosa | -0.16 | -0.09 | 0.359 | -2.92 | -3.19 |
| 83 | 50 | 4.2 | VidaRosa | 0.11 | -0.21 | 0.352 | -0.05 | -0.16 |
| 84 | 50 | 4.2 | VidaRosa | -0.33 | -0.35 | 0.452 | -0.12 | 0.11 |
| 85 | 50 | 3.8 | VidaRosa | -2.14 | -3.12 | 0.256 | -0.21 | -0.33 |
| 86 | 50 | 3.85 | VidaRosa | 0.64 | 1.58 | 0.198 | -1.84 | -2.14 |
| 87 | 50 | 3.85 | PDMS | 1.64 | 2.27 | 0.086 | 0.94 | 0.64 |
| 88 | 50 | 3.9 | PDMS | 2.23 | 3.13 | 0.341 | 1.62 | 1.64 |
| 89 | 50 | 4.1 | PDMS | 2.22 | 2.95 | 0.346 | 2.23 | 2.23 |
| 90 | 50 | 3.95 | PDMS | 2.34 | 3.15 | 0.149 | 2.10 | 2.22 |
| 91 | 50 | 5.4 | PDMS | -2.61 | -3.62 | 0.229 | 2.25 | 2.34 |
| 92 | 50 | 4.15 | PDMS | 2.82 | 3.83 | 0.166 | -2.58 | -2.61 |

Figure S4 shows a histogram of the standard deviation in spherical power of the fabricated lenses.

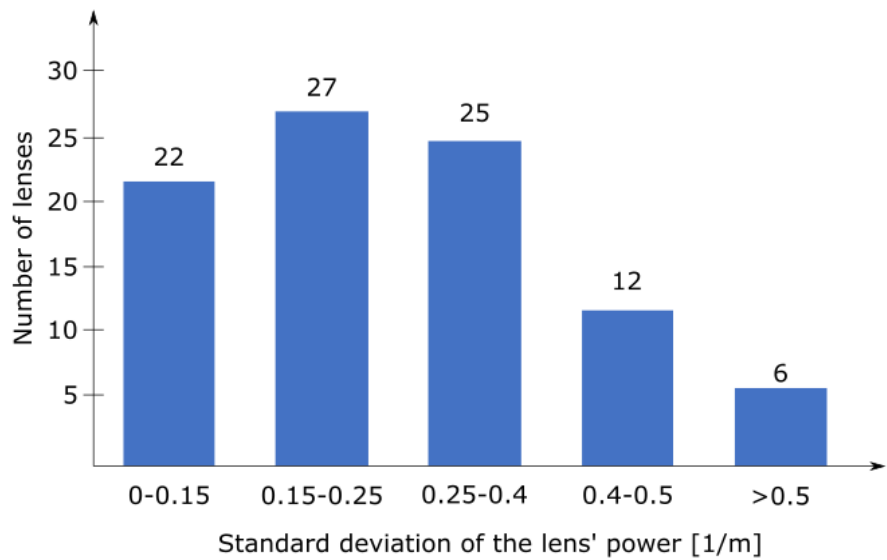

*Figure S4* – *A histogram of the standard deviation in power over the area of the fabricated lenses.*

Cylindrical lenses:

To validate the theory for the cylindrical power of a lens with elliptical boundaries, we fabricate lenses with different eccentricities and different volumes. Figure S5 shows the measured (symbols) and theoretical (solid lines) cylindrical power as a function of the total volume. Adjusting the frame eccentricity and the total liquid volume allows to produce any combination of cylindrical and spherical powers. For example, Figure S5(b) shows measurements of two lenses that have approximately the same cylindrical power (1.12 D, 1.08 D), and yet have different spherical powers - one has a positive diopter of 3.2 D and the other has a negative diopter of -0.78 D. For both lenses, the standard deviation in power, relative to the expected theoretical value, is smaller than 0.2 D. Table S2 lists the fabricated elliptical lenses, the frame geometries, and the measured spherical and cylindrical powers.

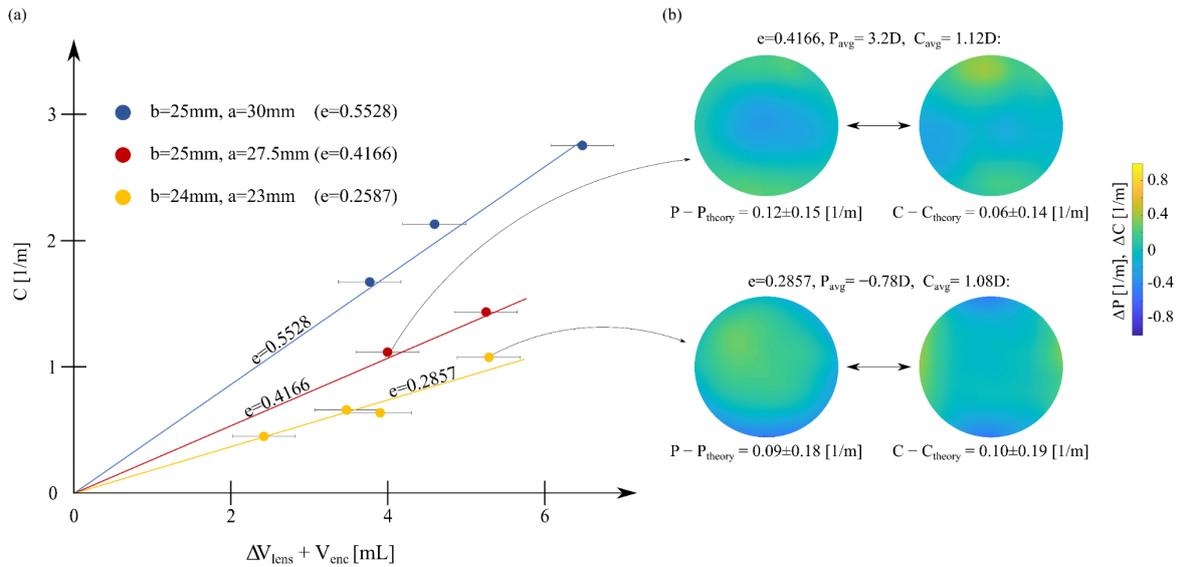

***Figure S5 – Experimental results and characterization of cylindrical lenses fabricated using the Fluidic Shaping method.*** *(a) The cylindrical power as a function of the injected volumes $\Delta V_{lens} + V_{enc}$, for 9 lenses with three different eccentricities. The refractive index of all lenses was $n = 1.525$. The solid lines represent the theoretical prediction according to Equation (27), showing good agreement with the experimental results. (b) The deviation of the measured spherical and cylindrical diopter maps $P(x,y), C(x,y)$ from their corresponding theoretical values $P_{theory}(x,y), C_{theory}(x,y)$ for lenses that were fabricated on different frames, showing the ability to decouple the powers. The results agree with the theoretical values and show standard deviations smaller than 0.2 diopters.*

***Table S2*** *– Fabricate elliptical-frame lenses, measured mean powers and standard deviation, and the theoretical values of the optical powers*

| | Bottom diameter $2R_0$ [mm] | Top small diameter $2b$ [mm] | Top large diameter $2a$ [mm] | Width $t$ [mm] | $\Delta V_{lens} = V_{lens} - V_{frame}$ [mL] | $V_{enc}$ [mL] | Measured P [1/m] | P STD [1/m] | Measured C [1/m] | C STD [1/m] | Theoretical P [1/m] | Theoretical C [1/m] |
|---|---|---|---|---|---|---|---|---|---|---|---|---|
| 1 | 50 | 50 | 55 | 3 | 1.99 | 3.25 | 1.77 | 0.104 | 1.43 | 0.140 | 1.86 | 1.40 |
| 2 | 50 | 50 | 55 | 3 | 2.50 | 1.48 | 3.20 | 0.216 | 1.11 | 0.195 | 3.09 | 1.06 |
| 3 | 50 | 50 | 60 | 5 | 1.07 | 3.52 | -0.55 | 0.113 | 2.13 | 0.178 | -0.47 | 1.98 |
| 4 | 46 | 46 | 48 | 4.85 | 0.77 | 4.51 | -0.78 | 0.101 | 1.07 | 0.084 | -0.68 | 0.97 |
| 5 | 46 | 46 | 48 | 4.8 | 0.49 | 3.40 | -0.48 | 0.135 | 0.639 | 0.295 | -0.71 | 0.72 |
| 6 | 46 | 46 | 48 | 4.85 | 1.34 | 1.07 | 2.65 | 0.096 | 0.452 | 0.147 | 2.34 | 0.44 |
| 7 | 46 | 46 | 48 | 4.8 | 0.62 | 2.84 | 0.07 | 0.108 | 0.660 | 0.158 | -0.13 | 0.64 |
| 8 | 50 | 50 | 60 | 4.8 | 0.22 | 6.25 | -2.42 | 0.388 | 2.75 | 0.531 | -2.84 | 2.79 |
| 9 | 50 | 50 | 60 | 4.9 | 0.15 | 3.61 | -1.69 | 0.107 | 1.67 | 0.189 | -1.61 | 1.62 |

## SI 7. AFM measurements

To assess the surface quality of lenses fabricated in our method and determine their suitability for use as eyewear lenses, we conducted tests using an Atomic Force Microscope (MFP-3D Infinity, Asylum Ltd by Oxford Instruments, UK). We analyzed nine locations on four different lenses. The results show a root-mean-squared (RMS) surface roughness of 1.4 nm, with individual measurements ranging from 0.43 nm to 3.3 nm. The table below lists the lenses measured, areas tested, and the surface RMS roughness.

*Table S3 – AFM measurements of lenses fabricated using the Fluidic Shaping method*

| # Measurement | Lens type | Area [μm X μm] | RMS value [pm] |
|---|---|---|---|
| 1 | Circular frame, P= –2.3D | 3X0.5 | 602 |
| 2 | Circular frame, P= 1.1D | 3X3 | 1185 |
| 3 |  | 3X3 | 3292 |
| 4 | Circular frame, P= –4D | 3X3 | 430 |
| 5 |  | 3X3 | 1110 |
| 6 |  | 3X3 | 2994 |
| 7 | Elliptical frame, P= –1.7D C= 1.6D | 1.5X1.5 | 631 |
| 8 |  | 1.5X1.5 | 992 |
| 9 |  | 3X3 | 1581 |